\documentclass[referee]{aa} 
\usepackage{graphicx}
\usepackage{natbib}
\usepackage{txfonts}
\def\hh{H$_{\rm 2}$}
\def\dd{D$_{\rm 2}$ }

\def\hdp{D$^+$ }
\def\hhhp{H$_3^+$ }
\def\hhdp{H$_2$D$^+$}

\newcommand   {\pscm}  {\mbox{cm$^{-2}$}}

\def\sm{s$^{-1}$}
\def\scm{cm$^{-2}$}
\def\ccm{cm$^{-3}$}
\def\cpm{cm$^{3}$}

\begin{document}
\title{Incorporation of stochastic chemistry on dust grains  
in the PDR code using moment equations}

\subtitle{I. Application to the formation of  \hh ~ and HD}

\author{F. Le Petit
\inst{1}
\and
B. Barzel\inst{2}
\and
O. Biham\inst{2}
\and
E. Roueff\inst{1}
\and
J. le Bourlot\inst{1}
}

\institute{Observatoire de Paris, LUTH 
and Universit\'e Denis Diderot, 
Place J. Janssen, 92190 Meudon, France\\
\email{franck.lepetit@obspm.fr}
\and
Racah Institute of Physics, The Hebrew University, 
Jerusalem 91904, Israel\\
}

\date{Received September 15, 2008; accepted }

 
\abstract
{Unlike gas-phase reactions, chemical reactions taking place 
on interstellar dust grain surfaces cannot always be modeled
by rate equations. Due to the small grain sizes and low flux,
these reactions may exhibit large fluctuations and thus require 
stochastic methods such as the moment equations. 
}
{We evaluate the formation rates of \hh, HD  and \dd   
molecules on dust grain surfaces 
and their abundances in the gas phase
under interstellar conditions.
}
{We incorporate the moment equations into the 
Meudon PDR code and compare the results with those
obtained from the rate equations.
}
{We find that within the experimental constraints on 
the energy barriers for diffusion and desorption and for the
density of adsorption sites on the grain surface,
\hh, HD and \dd molecules can be formed efficiently on dust grains.
}
{ 
Under a broad range of conditions, the moment equation
results coincide with those obtained from the rate equations.
However, in a range of relatively high grain temperatures, there
are significant deviations.
In this range, the rate equations fail while the moment equations
provide accurate results.
The incorporation of the moment equations into the 
PDR code can be extended to other reactions taking place on
grain surfaces.
}


\maketitle

%

\section{Introduction}

Most of the detected interstellar molecules are in the gas phase.
The chemical balance of gas phase species in interstellar clouds
is commonly described by a set of rate equations, using 
the known rate coefficients of gas-phase reactions.
Many studies have been devoted to 
the resolution of the chemical rate equations 
to derive the compositions of interstellar clouds within 
specific physical conditions. 
Two main approaches are used by modellers:
the time-dependent approach and the steady-state approach.
In the time-dependent approach, one 
follows the time evolution of the abundances of various atoms and 
molecules by numerically integrating the corresponding set of
first order coupled differential 
equations, from specific initial conditions. 
In the steady state approach, 
one directly solves the set of algebraic equations
which is obtained when the time derivatives of the
abundances are set to zero.
Comparison with observations allows then to 
test the relevance of the various hypotheses.
However, some important reactions which give rise 
to the formation of molecular hydrogen, ice mantles
and certain organic molecules do not take place in
the gas phase, but on the surfaces of dust grains.
To account for these surface processes, one needs to 
incorporate the grain-surface reactions into the gas 
phase models of interstellar chemistry. 
It is thus attractive to use rate equations for the surface reactions, 
which enable to couple surface reactions to gas-phase 
reactions in a straightforward way. 
This approach is indeed valid in the case of sufficiently 
large grains, when the number of reactive atoms and 
molecules of each species on a grain is sizeable. 
However, in the limit of small grains, 
when the number of reactive atoms and molecules on a grain is small, 
the rate equations are not always suitable. 
Indeed, rate equations simply ignore the fluctuations 
in the number of reactive species on the grains. 
This problem was 
discussed by several authors over many years 
\citep{tielens:82,char:97,caselli:98,shalab:98,stantcheva:01,stantcheva:02}.
To overcome these difficulties, 
a master equation approach has been proposed 
and was specifically applied to the formation of 
molecular hydrogen.
This approach is suitable for the simulation of diffusive 
chemical reactions on interstellar grains  
in the accretion limit 
\citep{biham:01,green:01,biham:02}. 
The master equation approach takes into account 
the discrete nature of reactive species on the surfaces
of grains as well as the fluctuations in their populations. 
This approach was later used for a larger chemical network
on grains, leading to the formation of methanol
and its deuterated versions
\citep{stantcheva:02,stantcheva:03}.
However, the number of equations included in the master
equation increases exponentially with the number of species
that are reactive on grain surfaces.
Therefore, it is not suitable for incorporation in codes
that include complex networks of grain-surface reactions. 
In addition to the master equation method,  
a Monte Carlo method was proposed by 
\cite{tielens:82} 
and applied to interstellar modelling by 
\cite{charnley:01}, 
\cite{caselli:02} 
and 
\cite{stantcheva:03}. 
Recently, 
\cite{cuppen:05}
developed random walk models for molecular hydrogen formation
on grains that take into account the effect of surface roughness
on the diffusion and reaction rates. 
They simulated these models using Monte Carlo methods
and showed that surface roughness tends to broaden the
range of temperatures in which molecular hydrogen formation
is efficient.
The advantage of the random walk models is that they enable
to account in more detail for the microphysics of the grain
surface. In particular, they enable to take into account the
whole distribution of binding energies and diffusion barriers
for H atoms. Detailed models of this type provide useful insight
but it is not feasible to include them in large models of
interstellar chemistry due to their complexity.
In contrast, the rate equation and moment
equation methods used in this paper, include a single energy 
barrier for each process. 
Finally, a semi-empirical approach, 
known as the modified rates method, 
was proposed by 
\cite{caselli:98}. 
This method is easy to employ and has 
been used with mixed success by
\cite{shalab:98}, 
\cite{stantcheva:02}
and 
\cite{caselli:02}.
\cite{Garrod:08} has studied a different version of the modified rate method.

The recently proposed moment equations method 
provides efficient stochastic simulations 
of grain-surface chemistry  
\citep{barzel:07a,barzel:07}. 
The method consists of only one equation for the average population 
size of each reactive species on a grain and one equation for
the rate of each reaction.
It consists of a set of coupled ordinary differential
equations, that resemble the rate equations.
Therefore, they can be easily coupled to the rate equations
of gas-phase chemistry.
Moreover, unlike the rate equations, 
the moment equations are linear, 
and thus easier to handle both in the time-dependent and
in the steady state approaches.
The moment equations were tested for the reaction network
that gives rise to ice mantles on grains,
that consist of water ice, carbon dioxide and methanol
\citep{barzel:07a,barzel:07}.
The stability properties of the moment equations
as well as the accuracy of their
steady-state solution 
under astrophysically relevant conditions
were examined by 
extensive computer simulations and by comparison 
to the master equation results.

In this paper we incorporate the moment 
equations into the Meudon PDR code 
\citep{lepetit:02,lepetit:06}, 
a stationary model of Photon Dominated Regions.
%
%
The model considers a stationary plane-parallel slab 
of gas and dust illuminated by an ultraviolet radiation 
field coming from one or both sides of the cloud. 
It solves, in an iterative way, at each point in the cloud, 
the radiative transfer in the UV, taking into account the absorption 
in the continuum by dust and in discrete transitions of H and H$_2$. 
Explicit treatment is performed for C and S photoionization, H$_2$ 
and HD photodissociation, as well as CO (and its isotopomeres) 
predissociating lines. The model also computes the thermal balance, 
taking into account heating processes such as the photoelectric 
effect on dust, cosmic rays and chemistry.  
It also accounts for the
cooling due to infrared and millimeter emission 
of the abundant ions, atoms, and molecules. 
The chemistry is solved under steady state conditions 
and the abundance of each species is computed 
at each point. 
The excitation states of a few important 
species are then computed. 
The column densities of these chemical species 
and their emissivities/intensities are then calculated.
To examine the applicability and relevance
of the moment equations within the PDR code, 
we consider a simple network of grain surface chemistry, 
involving only H, D, \hh, HD and D$_2$.
We compare the results obtained from the PDR code with
the moment equations with those obtained when the
rate equations are incorporated in the same code.
Unlike previous studies of molecular hydrogen formation that 
assumed a single grain size, the moment equations enable us
to take into account the full distribution of grain sizes.
We show that in case of
relatively high grain 
temperatures, the rate equations are not suitable and 
the moment equations should be used.
We also demonstrate the importance of the Langmuir
rejection effect at very low grain temperatures, where
the grain surfaces are saturated by hydrogen atoms
and reaction rates are low.

The paper is organized as follows. In section \ref{sec:chemistry} we briefly review the gas-phase and surface reactions involved in 
the formation of \hh, HD and \dd molecules. We present in section \ref{sec:moment} the moment 
equations for the H-D system and describe their incorporation into the PDR model. Simulation results are displayed in 
section \ref{sec:result}. The results are summarized and discussed in section \ref{sec:conclusion}.   
 

\section{Formation of Molecular Hydrogen and its Deuterated Versions}
\label{sec:chemistry}

\subsection{Astrophysical context}

No efficient gas-phase mechanism is at work to 
form molecular hydrogen in the gas phase.  
Therefore, reactions on grain surfaces 
are needed in order to account for the observed abundance 
of \hh. 
\cite{hollenbach:71} have been among the first to 
quantitatively describe the formation of molecular 
hydrogen on the surfaces of spherical dust grains.
Here we assume that
the grain size distribution follows a power law of the form

\begin{equation}
n(a) = c a^{-q}, 
\end{equation}

\noindent
as suggested by \citet*{mathis:77}.
The prefactor $c$ is given by

\begin{equation}
c = \frac{3}{4\pi} \frac{1.4 \cdot m_{\rm H} \cdot G \cdot (4-q)}
{\rho (a_{\rm max}^{4-q} - a_{\rm min}^{4-q})}n_{\rm H}
\end{equation}
\noindent
where $\rho$ (g cm$^{-3}$) stands for the volumic mass of the grains,
while $a_{\rm max}$ and $a_{\rm min}$ are the upper and lower cutoffs
of the grain radii distribution, respectively.  We denote
the proton density in the gas phase by $n_{\rm H}$ (\ccm) 
and the dust to gas mass ratio by $G=0.01$. 
In this case, one can formulate the formation 
rate of H$_2$ as a function of various  grain parameters such as the 
dust to gas mass ratio, the minimum and maximum values 
of the grain radii and the volumic mass of the grains. 
If the exponent of the power law is $q=3.5$, 
analytical formulae are obtained as derived by 
\citet{lebourlot:95}.
\cite{lepetit:02} 
have specifically studied 
the D/HD transition occuring at the edge of a 
translucent cloud and in dense photon dominated regions. 
They have also studied the formation of HD molecules on the surfaces 
of dust grains, following the procedure of 
\cite{lebourlot:95}. 
The resulting formulae are very similar to the pure molecular 
hydrogen case. 
Following previous considerations by 
\cite{watson:74} 
and 
\citet*{dalgarno:73}, 
\cite{lepetit:02} 
found that HD is much more efficiently formed in the gas phase, 
in a succession of reactions initiated by cosmic 
ray ionization of atomic hydrogen. 
When protons are formed,
the sequence of reactions involved is the following:

\begin{equation}
{\rm H^{+} +  D \rightleftharpoons D^{+} + H.}
\end{equation}

\noindent
The forward reaction is exothermic by an energy 
of 3.7 meV, corresponding to 43~K, 
so that the reverse reaction may take place at 
moderate temperatures, 
reducing the \hdp formation efficiency. 
\hdp reacts efficiently with \hh ~to form HD:

 \begin{equation}
{\rm D^{+} +  H_2 \rightarrow HD + H^{+}.}
\end{equation}

\noindent
The reverse reaction may also occur, 
but the endothermic barrier is now 40 meV, 
equivalent to 464~K, so that the corresponding 
probability is negligible at low temperatures. 
The main destruction channel of HD in diffuse 
environments is photodissociation, which has been computed by 
\cite{lepetit:02}. 
This simple gas-phase scheme accounts for the 
observed HD/H$_2$  abundance ratio in diffuse 
clouds and translucent regions, 
as found from Copernicus and FUSE observations 
\citep{watson:74,dalgarno:73,lacour:05}. 
In dark and cold regions, HD is the major repository of deuterium. 
It reacts with \hhhp to form \hhdp which
is the starting point of a rich chemistry of deuterium fractionation 
\citep{millar:89,roueff:00,caselli:02,stantcheva:03,roberts:03,roueff:05}. 
Although the formation of HD in diffuse clouds occurs primarily in the gas, 
it was recently proposed that under certain conditions, 
grain-surface reactions may also contribute to an enhanced 
production of HD and \dd molecules. 
The proposed mechanism is based on the assumption 
that D atoms stick more strongly than H atoms so 
that their desorption rate is lower. Such an isotope 
effect has been observed in various experimental situations 
\citep{koehler:88,hoogers:95,amiaud:07}. 
As a result, the residence time of D atoms on grains is 
expected to be longer than that of H atoms. 
Then, the D/H abundance ratio on the grains is enhanced 
compared with the gas-phase ratio and newly adsorbed 
H (or D) atoms are more likely to find D atoms already 
residing on the grains. This may give rise to an enhanced 
production of HD molecules. 

\subsection{The interaction between hydrogen/deuterium atoms and dust grains}

In order to quantitatively describe the formation 
of \hh, HD and \dd molecules on grain surfaces, 
one has to introduce several hypotheses and parameters, 
which we recall below. The typical velocities of H and D 
atoms in the gas phase are given by 
$\upsilon_{\rm H}$ 
and 
$\upsilon_{\rm D}$, 
respectively.
Under the simplifying assumption that all
grains are of the same radius, $a$, 
one can compute the numerical density of spherical dust grains 
$n_{\rm gr}$ (\ccm)
as a function of $n_{\rm H}$:

\begin{equation}
n_{\rm gr} =1.4 \cdot \frac{3  \cdot m_{\rm H} \cdot G}
                           {4\pi \cdot a^3 \cdot \rho }
                      \cdot n_{\rm H}.
\end{equation}

\noindent
For a power law distribution of grain sizes with the 
exponent $q=3.5$ \citep{mathis:77}, 
the numerical density of the grains becomes:

\begin{equation}
n_{\rm gr} = 
1.4 \cdot \frac{3  \cdot m_{\rm H} \cdot G 
\cdot (a_{\rm max}^{2.5}-a_{\rm  min}^{2.5})}
{ 20 \pi \cdot \rho \cdot (a_{\rm min}a_{\rm max})^{2.5} 
\cdot (\sqrt{a_{\rm max}} - \sqrt{a_{\rm min}}) } \cdot n_{\rm H},
\end{equation}

\noindent
The number of adsorption sites on a grain is denoted by $S$. 
Their density on the grain surface, $s$ (sites \scm), 
is given by 
$s = S / 4 \pi a^2$. 
The density of adsorption sites, $s$, 
and the distance between adjacent sites, 
$d$, are related through $d^2 = 1/s$. The fluxes 
$F_{\rm H}$ 
and 
$F_{\rm D}$ 
(atoms \sm) 
of H and D atoms onto the surface 
of a single grain are given by
$F_{\rm I} = n_{\rm I} \upsilon_{\rm I} \sigma$,   
where 
I = H, D, respectively and $\sigma$ is the 
cross section of a grain,
namely 
$\sigma = \pi a^2$. 

%
%

The atoms stick to the surface and hop as random 
walkers until they either desorb or recombine into molecules. 
The desorption rates of H and D atoms on the surface are given by

\begin{equation}
W_{\rm I} = 
\nu_{\rm I} \cdot 
\exp\left( - \frac{E_{\rm I}^{\rm des}}{k T_{\rm gr}} \right), 
\end{equation}

\noindent
where $\nu_{\rm I}$ 
is the attempt rate, $E_{\rm I}^{\rm des}$
is the energy needed for desorption of an atom of isotope 
I and $T_{\rm gr}$ is the grain temperature.
The hopping rates of the atoms on the surface are 

\begin{equation}
a_{\rm I} = 
\nu_{\rm I} \cdot 
\exp\left( - \frac{E_{\rm I}^{\rm diff}}{k T_{\rm gr}} \right), 
\end{equation}

\noindent
where 
E$_{\rm I}^{\rm diff}$ 
is the activation energy barrier for diffusion of the isotope I. 
The rate $a_{\rm I}$ is the inverse of the residence time
$\tau_{\rm I}$ of an atom of isotope I in a single adsorption site.
The sweeping rate $A_{\rm I} = a_{\rm I }/ S$ is
approximately the inverse of the time $S \tau_{\rm I}$ 
required for an atom of isotope I to visit nearly all 
the adsorption sites on the grain surface. 
Since the D atom is twice as heavy as the H atom, its ground
state energy within an adsorption site on the surface is lower.
Due to this isotope effect, 
the desorption barrier for D atoms is assumed to be
higher by 5 meV than for H atoms. 
For the diffusion barriers,
we assume 
$E_{\rm D}^{\rm diff} = E_{\rm H}^{\rm diff}$ 
because the diffusion barrier balances the zero point energy 
of the potential
well with that of the saddle point
(or transition state), 
while desorption does not possess such saddle point. 
Note that in the analysis of the experimental results,
presented in 
\cite{Katz:99}
and
\cite{Perets:05},
no distinction was made between H and D atoms,
in order to keep the number of fitting parameters small.

The possibility of tunneling 
of the adsorbed atoms between adsorption sites
has recently been studied by
\cite{cazaux:04}.

The tunneling 
probability depends on the distance between adjacent sites 
and on the energy barrier, 
which is taken as the diffusion barrier.
For the sake of completeness, we present
the corresponding tunneling rate,
which is given by
\citep{eisberg:61} :

\begin{equation}
k_{\rm tun} = {\nu_{\rm I}}
\left[ 
{1+\frac{(E_{\rm I}^{\rm diff})^2}
{4{\rm E}~(E_{\rm I}^{\rm diff}-{\rm E})} 
\cdot {\rm sinh^2}(d/ \lambda_{\rm DB})}
\right]^{-1},
\end{equation}

%
%

\noindent
where
$\lambda_{\rm DB}$ corresponds to the De Broglie wavelength 
and is given by  

\begin{equation}
\lambda_{\rm DB} = \frac{\hbar}{\sqrt{2m(E_{\rm I}^{\rm diff}-E)}}, 
\end{equation}

\noindent
and $E$ is the kinetic energy of the adsorbed atom, 
determined by the grain temperature, 
namely $E = k_B T_{\rm gr}$. 
The efficiency of the tunnelling process is sizeable 
only when the ratio
$d/\lambda_{\rm DB}$ 
is of order 1 or less.
There is no specific reason to assume that the adsorption sites 
are uniformly distributed. However, for simplicity we assume
that the distance $d$ between adjacent sites 
is fixed and given by
$d^2 = 1/s$, as explained above.
We have evaluated the tunneling rate for the parameters used
in this paper, and it was found to be negligible.
Therefore, in the simulation results presented in this
paper, the mobility of H and D atoms on grains is only
due to thermal hopping, while tunneling is ignored.

In Table \ref{tab:grain} 
we display the parameters 
that describe the interaction of H and D atoms
with grains of different compositions and surface morphologies,
as reported in the literature. 
These parameters are based on a series of experiments
and subsequent analysis,
reported in \cite{Katz:99}, \cite{Perets:05} and \cite{Perets:07}.
The attempt rate $\nu_{\rm I}$ is often assumed 
to be equal to 10$^{12}$ \sm. 
It may also be derived from an harmonic 
oscillator model, where 
$\nu_{\rm I} = \sqrt{2E^{\rm des}_{\rm I}/m} ~ /  ({\pi ~d}) $.


\begin{table*}[!t]
\centering
\caption[]{Parameters for the interaction of hydrogen
and deuterium atoms with dust grains of different 
compositions and surface morphologies. 
Numbers in parentheses refer to powers of ten.
}
\label{tab:grain}
$$ 
\begin{tabular}{|l||l|l|l|l|}
\hline
\noalign{\smallskip}
& Amorphous carbon &   Olivine  &  Amorphous silicate & Low density ice   \\
Parameter & Katz (1999) &  Katz (1999) &  Perets (2007) & Perets (2005) \\
\noalign{\smallskip}
\hline
\hline
\noalign{\smallskip}
$\rho$ (g \ccm)  &  2.16 &  $3^a$  & 3.5 & 0.94\\
$s$ (\scm) & $5(13)$ &  $2(14)$ & $7(14)$ & 5(13)$^a$\\
$d$ ($\AA$)  &  14.14 &  7.07  & 3.78 &  14.14$^a$\\
$E_{\rm H}^{\rm des}$ (meV)  &  56.7 &  32.1 & 44.0 & 52.3 \\
$E_{\rm D}^{\rm des}$ (meV)  &  61.7 &  37.1 & 49.0 & 57.3 \\
$E^{\rm diff}_{\rm I}$ (meV)  &  44.0 &  24.7 & 35.0 & 44.5\\
$\nu_{\rm I}$ & $10^{12}$ & $10^{12}$ & $10^{12}$ & $10^{12}$ \\
\noalign{\smallskip}
\hline
\end{tabular}
$$ 
\begin{list}{}{}
\item[$^{\mathrm{a}}$] 
Assumed sensible values, as they are not given in the cited papers.
\end{list}
\end{table*}

The reported values of the desorption and diffusion barriers 
put severe constraints on the grain temperature range over which 
molecular hydrogen formation may occur. If one considers a distribution 
of grains of various sizes, such as the power law derived by 
\cite{mathis:77}, 
one has to perform the average of the various quantities over 
the size distribution, keeping in mind that different temperatures 
may pertain to grains of different sizes. In the present paper, we  
assume that the grain temperature depends on the surrounding radiation
field but not on the grain size. Under this assumption, in a given 
location in the cloud, grains 
of all sizes have the same temperature.
In the simulations reported in this paper we use the parameters of
amorphous carbon. Amorphous carbon is assumed to be a primary
component of interstellar dust. 
Its surface properties are thus suitable for the simulation of 
molecular hydrogen formation in relatively warm 
regions in which the grain surfaces are not covered by ice mantles.
It is also observed that the parameter values of amorphous carbon are 
close to those of low density amorphous ice, reported by \cite{Perets:05}.
Therefore, we use the same set of parameters for the interaction of 
hydrogen atoms with grain surfaces, 
independently of whether these grains are expected to be bare
or covered by ice mantles.

%

\section{Rate equations, master equation and moment equations}
\label{sec:moment}

\subsection{Rate equations}

After being adsorbed onto the surface, 
the H and D atoms hop between adsorption 
sites until they either desorb or 
form new molecules. The number of atoms of isotope I 
on the surface of a grain is denoted by 
$N_{\rm I:}$. 
The rate equations account for the expectation values 
$\langle N_{\rm I:} \rangle$, 
of the population size of isotope I on a grain
of a given radius where
I = H or D. 

\subsubsection{Basic processes}

In the present version of the PDR code, 
we introduce three main physical processes 
to describe the formation of  
H$_2$, HD and D$_2$ on grain surfaces:
adsorption, diffusion-mediated reaction and desorption:

\begin{itemize}

\item  {\bf {Adsorption}  }

\begin{tabular}{llllll}
H &+& grain &$\rightarrow$ & H:   & ($k^{\rm H}_{\rm ads}$ in s$^{-1}$) \\ 
D &+& grain &$\rightarrow$ & D:   & ($k^{\rm D}_{\rm ads}$ in s$^{-1}$),  
\end{tabular}
\\
\\
where
H: and D: stand for the hydrogen and deuterium atoms
which are adsorbed on the grain surface.
The corresponding rate equations are:

\begin{equation}
{\frac{dn({\rm H:})}{dt} = +k^H_{\rm ads} n(H)}
\end{equation}
\begin{equation}
{\frac{dn({\rm D:})}{dt} = +k^D_{\rm ads} n(D)}.
\end{equation}

\noindent
The adsorption coefficients 
$k^H_{\rm ads}$ 
and
$k^D_{\rm ads}$ 
are directly  proportional to  the grain cross section 
$\sigma = \pi a^2$ 
and to the sticking probability
$\gamma$, namely 

\begin{equation}
k^{\rm H}_{\rm ads} =  
\frac{F_{\rm H}}{n(H)} n_{\rm gr} 
= \gamma \upsilon_{\rm H} \pi a^2 n_{\rm gr}.  
\end{equation}

\noindent
For a distribution of grain sizes, one has to perform 
the average on 
$\sigma n_{\rm gr}$ 
over the distribution function, 
which leads to the following expression:

\begin{equation}
k^{\rm H}_{\rm ads} =   \langle \frac{F_{\rm H}}{n(H)} 
n_{\rm gr} \rangle=  \gamma \cdot \upsilon_{\rm H} \cdot 1.4 
\cdot \frac{3  \cdot m_{\rm H} \cdot G}{4\pi 
\cdot \rho \sqrt{a_{min} a_{max}} } \cdot  n_{\rm H}  
\end{equation}

\noindent
Similar equations hold for deuterium adsorption.\\

\item  
{\bf {Reaction on the grain surface due to diffusion}}

The diffusion-mediated formation of molecular hydrogen
and its deuterated versions is described by:

\begin{tabular}{lllllllll}
$\rm H:$ &   +&$\rm H: $ &$\rightarrow$ & $\rm H_2$  
        & & & &  (k$^{\rm H_2}_{\rm surf}$ in \ccm s$^{-1}$)   \\ 
$\rm H:$ &   +&$\rm D: $ &$\rightarrow$ & $\rm HD$ 
        & & & &  (k$^{\rm HD}_{\rm surf}$ in \ccm s$^{-1}$)  \\ 
$\rm D:$ &   +&$\rm D: $ &$\rightarrow$ & $\rm D_2$ 
         & & & & (k$^{\rm D_2}_{\rm surf}$ in \ccm s$^{-1}$).  \\ 
\end{tabular}

\noindent
Assuming that all formed molecules are directly released into the gas phase, the formation rates of 
H$_2$, HD and D$_2$ are given by :

\begin{equation}
{\frac{dn({\rm H_2})}{dt} = 
k^{\rm H_2}_{\rm surf} n(H:)^2} = 
A_H \langle N_{\rm H:} \rangle^2 n_{gr}
\end{equation}

\begin{equation}
{\frac{dn({\rm HD})}{dt} = 
k^{\rm HD}_{\rm surf} n(\rm H:)n(\rm D:)} = 
({A_H + A_D}) 
\langle N_{\rm H:}  \rangle  
\langle N_{\rm D:}  \rangle n_{gr}
\end{equation}

\begin{equation}
{\frac{dn({\rm D_2})}{dt} = 
k^{\rm D_2}_{\rm surf} n(D:)^2} 
= A_D \langle N_{\rm D:} \rangle^2 n_{gr},
\end{equation}

where 
$\langle N_{\rm H:} \rangle$ 
($\langle N_{\rm D:} \rangle$) 
is the number of adsorbed hydrogen (deuterium) atoms on a single grain.
The rate coefficients corresponding 
to these surface reactions can be obtained 
from the relations:
$n(H:) = \langle N_{\rm H:} \rangle \cdot n_{gr}$ 
and
$n(D:) = \langle N_D \rangle \cdot n_{gr}$.
These rate coefficients take the form

\begin{eqnarray}
k^{\rm H_2}_{\rm surf} &=& 
\frac{A_H \langle N_{\rm H:} \rangle^2}{n(H:)^2}n_{gr}=
\frac{A_{\rm H}}{n_{gr}} \nonumber \\
k^{\rm HD}_{\rm surf} &=&
\frac{{(A_{\rm H}+A_{\rm D})} \langle N_{ \rm H:} \rangle 
\langle N_{\rm D:} \rangle}{n(H:)n(D:)}n_{gr} =
\frac{A_{\rm H}+A_{\rm D}}{n_{gr}} \nonumber \\
k^{\rm D_2}_{\rm surf} &=&
\frac{A_{\rm D} \langle N_{\rm D:}\rangle^2}{n(\rm D:)^2}n_{gr}=
\frac{A_{\rm D}}{n_{gr}}. \nonumber 
\end{eqnarray}

\noindent
These formulae are valid only if the grains 
have the same temperature because the diffusion coefficient 
is temperature dependent. \\

\item {\bf {Desorption}}

The desorption processes 

\begin{tabular}{lllllllll}
H: &     &$\rightarrow$ & H            &+& grain 
& & & (k$^{\rm H}_{\rm des}$ in s$^{-1}$) \\ 
D: &    &$\rightarrow$ & D            &+& grain 
& & &  (k$^{\rm D}_{\rm des}$ in s$^{-1}$),  
\end{tabular}

\noindent
are described by the equations

\begin{eqnarray}
 \frac{dn(\rm H:)}{dt}  = -k^{\rm H}_{\rm des}~n(\rm H:)  \\
 \frac{dn(\rm D:)}{dt}  = -k^{\rm D}_{\rm des}~n(\rm D:),  
\end{eqnarray}

\noindent
where

\begin{eqnarray}
k^H_{des} &=& W_H \\  
k^D_{des} &=& W_D.  
\end{eqnarray}

\noindent
The overall variations of adsorbed H and D atoms per \cpm~ 
in the rate equation formalism are:

\begin{eqnarray}
\label{rate:H}
\frac{d n({\rm H:}) }{dt} &=& 
k^H_{\rm ads}n(H) -W_{\rm H} n({\rm H:}) 
- 2\frac{A_{\rm H}}{n_{gr}} n({\rm H:}) ^2
- \frac{(A_{\rm H}+A_{\rm D})}{n_{gr}} n({\rm H:}) n({\rm D:}) \\ 
\label{rate:D}
\frac{d n({\rm D:}) }{dt} &=& 
k^D_{\rm ads}n(D) -W_{\rm D} n({\rm D:}) 
- 2\frac{A_{\rm D}}{n_{gr}} n({\rm D:}) ^2
- \frac{(A_{\rm H}+A_{\rm D})}{n_{gr}} n({\rm H:}) n({\rm D:})  
\end{eqnarray}

\end{itemize}

\subsubsection{Rejection effects}
 
As the number of adsorption sites on a grain is finite, 
one has to take into account the possibility that 
hydrogen atoms that impinge upon already occupied sites
may be rejected rather than adsorbed on the grain.
This mechanism is often referred to as Langmuir rejection. 
When the rejection is taken into account, 
the flux terms are modified according to

\begin{equation}
F^{\rm eff}_{\rm I}= F_{\rm I}(1 - \frac{ N_{\rm H} + N_{\rm D} }{S}).
\label{eq:rejterm}
\end{equation}

\noindent
where I stands for H or D.
Since the abundance of D atoms is very small (1.5 10$^{-5}$) compared
to H atoms, it is sensible to consider first, for simplicity, 
the case in which the rejection is only due to 
already adsorbed H atoms.

\begin{itemize}
\item  
{\bf Rejection only due to adsorbed H atoms}
 
We first consider the case of rejection only due to 
adsorbed hydrogen atoms.
The rate equation for adsorbed H on a single grain becomes:

\begin{equation}
\label{rate:1}
\frac{d \langle N_{\rm H:} \rangle}{dt} = 
F_{\rm H} 
- \left(W_{\rm H} 
+ \frac{F_{\rm H}}{S} \right) \langle N_{\rm H:} \rangle
- 2A_{\rm H} \langle N_{\rm H:} \rangle^2
- (A_{\rm H}+A_{\rm D}) \langle N_{\rm H:} \rangle
  \langle N_{\rm D:} \rangle. 
\end{equation}

\noindent
The net result is an apparent increase of 
the desorption term by 
$f_{\rm H} = {F_{\rm H}}/{S}$. 
When the incoming flux of deuterium atoms is similarly modified, 
the rate equation for the adsorbed deuterium becomes: 

\begin{equation}
\label{rate:2}
\frac{d \langle N_{\rm D:} \rangle}{dt} = 
F_{\rm D} 
- \left(W_{\rm H}
+ \frac{F_{\rm D}}{S}\frac{ \langle N_{\rm H:} \rangle}
  {\langle N_{\rm D:} \rangle} \right) \langle N_{\rm D:} \rangle
- 2A_{\rm D} \langle N_{\rm D:} \rangle^2
- (A_{\rm H}+A_{\rm D}) \langle N_{\rm H:} \rangle
  \langle N_{\rm D:} \rangle. 
\end{equation}

\noindent
The correction due to rejection introduces a 
term proportional to 
${\langle N_{\rm H:} \rangle}/{\langle N_{\rm D:} \rangle}$.
The corresponding equations for 
the quantities expressed in volumic density become the following:

\begin{equation}
\label{rate:H2}
\frac{d n({\rm H:}) }{dt} = 
k^H_{\rm ads}n(H) 
- {\left({W_{\rm H}+\frac{\gamma \upsilon_{\rm H} n(H)}{4s}}\right)} n({\rm H:}) 
- 2\frac{A_{\rm H}}{n_{gr}} n({\rm H:})^2
- \frac{(A_{\rm H}+A_{\rm D})}{n_{gr}} n({\rm H:}) n({\rm D:})  
\end{equation}
\begin{equation}
\label{rate:D2}
\frac{d n({\rm D:}) }{dt} =
k^D_{\rm ads}n(D) 
- \left(W_{\rm D} 
+ \frac{\gamma \upsilon_{\rm D} n(D)}{4s} \frac{n(H:)}{n(D:)}\right) n({\rm D:}) 
- 2\frac{A_{\rm D}}{n_{gr}} n({\rm D:}) ^2
- \frac{(A_{\rm H}+A_{\rm D})}{n_{gr}} n({\rm H:}) n({\rm D:})  
\end{equation}

\item{\bf Rejection by adsorbed H and D atoms}

We now include the adsorbed deuterium atoms in the rejection
term according to Eq.
(\ref{eq:rejterm}).
As the equations are very similar, 
we only display the resulting rate equations:

\begin{equation}
\label{rate:H3}
\frac{d n({\rm H:}) }{dt} = 
k^H_{\rm ads}n({\rm H}) 
- {W_{\rm H}^{\rm eff}} n({\rm H:}) 
- 2\frac{A_{\rm H}}{n_{\rm gr}} n({\rm H:})^2
- \frac{(A_{\rm H}+A_{\rm D})}{n_{\rm gr}} n({\rm H:}) n({\rm D:})  
\end{equation}

\begin{equation}
\label{rate:D3}
\frac{d n({\rm D:}) }{dt} = 
k^D_{\rm ads}n(D) 
- W_{\rm D}^{\rm eff} n({\rm D:}) 
- 2\frac{A_{\rm D}}{n_{\rm gr}} n({\rm D:})^2
- \frac{(A_{\rm H}+A_{\rm D})}{n_{\rm gr}} n({\rm H:}) n({\rm D:}).  
\end{equation}

\noindent
The values of the effective desorption coefficients are : 

\begin{equation}
\label{rate:H4}
W_{\rm H}^{\rm eff} =
{W_{\rm H}+\frac{\gamma \upsilon_{\rm H} n(H)}{4s}}
+ \frac{\gamma \upsilon_{\rm H} n(H)}{4s}\frac{n(D:)}{n(H:)}
\end{equation}
\begin{equation}
\label{rate:D4}
W_{\rm D}^{\rm eff} =
{W_{\rm D}+\frac{\gamma \upsilon_{\rm D} n(D)}{4s}}
+ \frac{\gamma \upsilon_{\rm D} n(D)}{4s}\frac{n(H:)}{n(D:)}.
\end{equation}

\end{itemize}
 
The production rates of H$_{\rm 2}$, HD and D$_{\rm 2}$ 
(in units of \ccm \sm) are: 

\begin{eqnarray}
R_{\rm H_2} &=& {(A_{\rm H }/n_{gr})} n({\rm H:})^2   \\ 
R_{\rm HD}  &=& {((A_{\rm H }+A_{\rm D})/n_{gr})}n({\rm H:})n({\rm D:}) \\ 
R_{\rm D_2} &=& {(A_{\rm D }/n_{gr})} n({\rm D:})^2.  
\end{eqnarray}

\noindent
This is a generalized form of the equations 
given in 
\cite{lipshtat:04}.

\subsection{Master and moment equations when only the  
rejection term for H atoms is included}

\subsubsection{Master equation}

When the number of adsorbed atoms is small,
the rate equations may become inaccurate.
In this case one may consider using the master equation
or the moment equations derived from it. 
The master equation describes the temporal 
evolution of the probabilities 
$P(N_{\rm H:}, N_{\rm D:})$ 
that 
N$_{\rm H:}$ hydrogen atoms
and 
N$_{\rm D:}$ deuterium atoms reside on
the surface of a given grain. 
Here we write the master equation 
when only the Langmuir rejection due to hydrogen atoms 
is taken into account. It takes the form 

\begin{eqnarray}
\frac{{d P}(N_{\rm H:}, N_{\rm D:})}{dt} &=& 
F_{\rm H}\left[\left(1-\frac{N_{\rm H:}-1}{S}\right)
P(N_{\rm H:}-1, N_{\rm D:}) -
\left(1-\frac{N_{\rm H:}}{S}\right)
P(N_{\rm H:}, N_{\rm D:})\right]  \nonumber \\  
&+ &F_{\rm D}\left[\left(1-\frac{N_{\rm H:}}{S}\right)
P(N_{\rm H:}, N_{\rm D:}-1) -
\left(1-\frac{N_{\rm H:}}{S}\right)
P(N_{\rm H:}, N_{\rm D:}\right]  \nonumber \\
&+&W_{\rm H} \left[(N_{\rm H:}+1)P(N_{\rm H:}+1, N_{\rm D:})
- N_{\rm H:}P(N_{\rm H:}, N_{\rm D:})\right]  \nonumber \\
& +& W_{\rm D} \left[(N_{\rm D:}+1)P(N_{\rm H:}, N_{\rm D:}+1)
- N_{\rm D:}P(N_{\rm H:}, N_{\rm D:})\right]  \nonumber \\
&+& A_{\rm H} \left[(N_{\rm H:}+2)(N_{\rm H:}+1)P(N_{\rm H:}+2, N_{\rm D:}) 
- N_{\rm H:}(N_{\rm H:}-1)P(N_{\rm H:}, N_{\rm D:})\right] \nonumber \\
& + &A_{\rm D} \left[(N_{\rm D:}+2)(N_{\rm D:}+1)P(N_{\rm H:}, N_{\rm D:}+2) 
- N_{\rm D:}(N_{\rm D:}-1)P(N_{\rm H:}, N_{\rm D:})\right] \nonumber \\
&+& {(A_{\rm H}+A_{\rm D}) \left[(N_{\rm H:}+1)(N_{\rm D:}+1)
P(N_{\rm H:}+1, N_{\rm D:}+1)
- N_{\rm H:} N_{\rm D:} P(N_{\rm H:}, N_{\rm D:}) \right]}.
\end{eqnarray}

\noindent
The set of equations is written for the various integer values of 
$N_{\rm H:}$ and $N_{\rm D:}$. 
These equations are essentially identical to 
those derived previously by 
\cite{lipshtat:04}, 
except for the fact that we have explicitely introduced the possibility
of the Langmuir rejections via the 
${-F \cdot N/S}$ terms.
Using the relation
$f_{\rm I} = F_{\rm I} /S$  
we rewrite the master equation in the following way:

\begin{eqnarray}
\frac{{d P}(N_{\rm H:}, N_{\rm D:})}{dt} &=& 
F_{\rm H}\left[P(N_{\rm H:}-1, N_{\rm D:})-P(N_{\rm H:}, N_{\rm D:})\right]
\nonumber \\
&+& F_{\rm D}\left[(P(N_{\rm H:}, N_{\rm D:}-1)
    -P(N_{\rm H:}, N_{\rm D:}\right]  \nonumber \\
&+& f_{\rm H} \left[N_{\rm H:}P(N_{\rm H:}, N_{\rm D:})
    - (N_{\rm H:}-1) P(N_{\rm H:}-1, N_{\rm D:})\right] \nonumber \\
&+& f_{\rm D}  \left[N_{\rm H:}P(N_{\rm H:}, N_{\rm D:})
    - N_{\rm H:} P(N_{\rm H:}, N_{\rm D:}-1)\right] \nonumber \\
&+& W_{\rm H} \left[(N_{\rm H:}+1)P(N_{\rm H:}+1, N_{\rm D:})
    - N_{\rm H:}P(N_{\rm H:}, N_{\rm D:})\right]  \nonumber \\
&+& W_{\rm D} \left[(N_{\rm D:}+1)P(N_{\rm H:}, N_{\rm D:}+1)
    - N_{\rm D:}P(N_{\rm H:}, N_{\rm D:})\right]  \nonumber \\
&+& A_{\rm H} \left[(N_{\rm H:}+2)(N_{\rm H:}+1)P(N_{\rm H:}+2, N_{\rm D:}) 
    - N_{\rm H:}(N_{\rm H:}-1)P(N_{\rm H:}, N_{\rm D:})\right] \nonumber \\
&+& A_{\rm D} \left[(N_{\rm D:}+2)(N_{\rm D:}+1)P(N_{\rm H:}, N_{\rm D:}+2) 
    - N_{\rm D:}(N_{\rm D:}-1)P(N_{\rm H:}, N_{\rm D:})\right] \nonumber \\
&+& {(A_{\rm H}+A_{\rm D}) 
    \left[(N_{\rm H:}+1)(N_{\rm D:}+1)P(N_{\rm H:}+1,N_{\rm D:}+1)
    - N_{\rm H:} N_{\rm D:} P(N_{\rm H:}, N_{\rm D:})         \right]}
\end{eqnarray}

\noindent
This set of equations is identical to the standard master equations system 
\citep{barzel:07} 
with two additional terms proportional 
to f$_{\rm H}$ and f$_{\rm D}$. 
By a suitable summation over the probabilities in 
the master equation, one can obtain the moments of 
the distribution of adsorbed species populations, 
defined by:

\begin{equation}
\langle N_{\rm H:}^kN_{\rm D:}^l \rangle = 
\sum_{N_{\rm H:},N_{\rm D:}} N_{\rm H:}^kN_{\rm D:}^l 
P(N_{\rm H:}, N_{\rm D:})
\end{equation}

\noindent
The order of each moment is defined by the sum $l+k$. 
The first-order moments 
$\langle N_{\rm H:} \rangle$ 
and 
$\langle N_{\rm D:} \rangle$ 
represent the mean number of adsorbed H and D atoms, 
respectively.

\subsubsection{Moment equations}

\cite{lipshtat:03} have shown that the moment equation 
formalism may describe adequately the evolution of the 
system and allows a significant reduction of the number 
of coupled equations that needs to be solved.
We apply the same technique as in 
\cite{barzel:07} to derive the corresponding 
moment equations when the additional terms introduced 
by the Langmuir rejection are included. 
The time derivatives of the moments introduced 
by the term proportional to $f_{\rm I}$ is 

\begin{eqnarray}
\frac{d\langle {N_{\rm H:}}  \rangle_{\rm f}}{dt} 
&=&   \sum_{N_H,N_D}  N_{\rm H:}  
\dot{P_{\rm f}}(N_{\rm H:}, N_{\rm D:}) 
=  - f_{\rm H} \langle N_{\rm H:} \rangle   \nonumber \\
\frac{d \langle {N_{\rm D:}}  \rangle_{\rm f}}{dt} &=&    
\sum_{N_H,N_D}  N_{\rm D:}  \dot{P_{\rm f}}(N_{\rm H:}, N_{\rm D:})
= - f_{\rm D} \langle N_{\rm H:} \rangle   \nonumber \\
\frac{d \langle {N_{\rm H:}}^2  \rangle_{\rm f}}{dt} &=&   
\sum_{N_H,N_D}  N_{\rm H:}^2  \dot{P_{\rm f}}(N_{\rm H:}, N_{\rm D:}) 
= - f_{\rm H} (\langle N_{\rm H:} \rangle 
+ 2 \langle N_{\rm H:}^2 \rangle )  \nonumber \\
\frac{d \langle {N_{\rm D:}}^2  \rangle_{\rm f}}{dt} &=&    
\sum_{N_H,N_D}  N_{\rm D:}^2  \dot{P_f}(N_{\rm H:}, N_{\rm D:}) 
= - f_{\rm D} (\langle N_{\rm H:} \rangle 
+ 2 \langle N_{\rm H:} N_{\rm D:} \rangle )  \nonumber \\
\frac{d \langle {N_{\rm H:}N_{\rm D:}}  \rangle_{\rm f}}{dt} &=&    
\sum_{N_H,N_D}  N_{\rm H:}  N_{\rm D:} \dot{P_f}(N_{\rm H:}, N_{\rm D:})
= -f_{\rm H} \langle N_{\rm H:}N_{\rm D:} \rangle
-f_{\rm D} \langle N_{\rm H:}^2 \rangle, 
\end{eqnarray}

\noindent
where the $f_{\rm I}$ dependent terms are given by

\begin{eqnarray}
\dot{P_{\rm f}}(N_{\rm H:}, N_{\rm D:}) &=& 
f_{\rm H} \left[ N_{\rm H:}P(N_{\rm H:},N_{\rm D:})
-(N_{\rm H:}-1)P(N_{\rm H:}-1,N_{\rm D:}) \right]   
\nonumber \\
&+& f_{\rm D} \left[ N_{\rm H:}P(N_{\rm H:},N_{\rm D:})
-N_{\rm H:}P(N_{\rm H:},N_{\rm D:}-1) \right]  
\end{eqnarray}

\noindent
Adding these terms to the standard moment equations, 
one obtains the following system:

\begin{eqnarray}
\frac{d \langle {N_{\rm H:}} \rangle}{dt} & =& 
F_{\rm H} + (2 A_{\rm H} 
- W_{\rm H}-f_{\rm H}) \langle N_{\rm H:} \rangle 
- 2 A_{\rm H} \langle N_{\rm H:} ^2 \rangle 
- (A_{\rm H}+ A_{\rm D}) \langle N_{\rm H:}N_{\rm D:} \rangle  \nonumber \\
\frac{d \langle {N_{\rm D:}} \rangle}{dt} &=& 
F_{\rm D}  -f_{\rm D} \langle N_{\rm H:} \rangle 
+ (2 A_{\rm D}- W_{\rm D})  \langle N_{\rm D:} \rangle
- 2A_{\rm D} \langle N_{\rm D:}^2 \rangle  
- (A_{\rm H}+ A_{\rm D}) \langle N_{\rm H:}N_{\rm D:} \rangle \nonumber \\
\frac{d \langle {N_{\rm H:}}^2 \rangle}{dt} &=&
F_{\rm H} 
+ (2 F_{\rm H} + W_{\rm H} 
+ 4 A_{\rm H} -f_{\rm H}) \langle N_{\rm H:} \rangle 
- (2 W_{\rm H} + 4 A_{\rm H}+2f_{\rm H} ) \langle N_{\rm H:} ^2 \rangle 
\nonumber \\
&-& (A_{\rm H} + A_{\rm D}) \langle N_{\rm H:}N_{\rm D:} \rangle  \nonumber \\
\frac{d \langle {N_{\rm D:}}^2 \rangle}{dt} &=&
F_{\rm D} 
- f_{\rm D} \langle N_{\rm H:} \rangle 
+ (2 F_{\rm D} + W_{\rm D} +4 A_{\rm D}) \langle N_{\rm D:} \rangle 
- (2 W_{\rm D} + 4 A_{\rm D}) \langle N_{\rm D:}^2 \rangle  
\nonumber \\
&-& (A_{\rm H} + A_{\rm D}+2f_{\rm D}) 
  \langle N_{\rm H:}N_{\rm D:} \rangle  \nonumber \\ 
\frac{d \langle {N_{\rm H:}N_{\rm D:}} \rangle}{dt} &=&  
F_{\rm D} \langle N_{\rm H:} \rangle 
+ F_{\rm H}  \langle N_{\rm D:} \rangle  
- f_{\rm D} \langle N_{\rm H:}^2 \rangle  
- ( W_{\rm H}+W_{\rm D}+ A_{\rm H} + A_{\rm D} + f_{\rm H}) 
  \langle N_{\rm H:}N_{\rm D:} \rangle   
\end{eqnarray}

\noindent
This is the set of equations that is actually used in the
interstellar chemistry model, in the case where rejection
only due to H atoms is included.
This system of coupled differential equations 
accounts for the populations of H and D atoms on a grain
(given by the two first order moments) 
and for the reaction rates
(given by the three second order moments). 
One then obtains the formation rates 
$r_{\rm H_2}$,
$r_{\rm HD}$
and
$r_{\rm D_2}$,
of \hh, HD and \dd,
respectively, 
on a single grain 
\citep{lipshtat:04}: 

\begin{eqnarray}
\label{equ:form}
r_{\rm H_2} &=&  
A_{\rm H} \left (\langle N_{\rm H}^2 \rangle 
- \langle N_{\rm H} \rangle \right) \nonumber \\
r_{\rm HD} &=& 
(A_{\rm H}+A_{\rm D} )
\langle N_{\rm H}N_{\rm D} \rangle  \nonumber \\
r_{\rm D_2} &=& 
A_{\rm D} \left ( \langle N_{\rm D} ^2 \rangle 
- \langle N_{\rm D} \rangle \right). 
\end{eqnarray}

\noindent
The formation rate of \hh, HD and \dd on grains is 
further obtained by summing over the number of grains.

\subsection{Master and moment equations with rejection terms 
for both H and D atoms}

When the Langmuir rejection terms due to both H and D atoms adsorbed on the surface are included, 
the master equation takes the form :

\begin{eqnarray}
\frac{{d P}(N_{\rm H:}, N_{\rm D:})}{dt} &=& 
F_{\rm H}\left[ \left(1-\frac{N_{\rm H:}-1}{S}
-\frac{N_{\rm D:}}{S}\right)P(N_{\rm H:}-1, N_{\rm D:}) -
\left(1-\frac{N_{\rm H:}}{S}-\frac{N_{\rm D:}}{S}\right)
P(N_{\rm H:}, N_{\rm D:})\right]   \nonumber \\ 
&+& F_{\rm D}\left[\left(1-\frac{N_{\rm H:}}{S}
-\frac{N_{\rm D:}-1}{S}\right)P(N_{\rm H:}, N_{\rm D:}-1) -
\left(1-\frac{N_{\rm H:}}{S}
-\frac{N_{\rm D:}}{S}\right)P(N_{\rm H:}, N_{\rm D:})\right] \nonumber \\
&+& W_{\rm H} \left[(N_{\rm H:}+1)P(N_{\rm H:}+1, N_{\rm D:})
- N_{\rm H:}P(N_{\rm H:}, N_{\rm D:})\right]   \nonumber \\
&+& W_{\rm D} \left[(N_{\rm D:}+1)P(N_{\rm H:}, N_{\rm D:}+1)
- N_{\rm D:}P(N_{\rm H:}, N_{\rm D:})\right]   \nonumber \\
&+& A_{\rm H} \left[(N_{\rm H:}+2)(N_{\rm H:}+1)P(N_{\rm H:}+2, N_{\rm D:}) 
- N_{\rm H:}(N_{\rm H:}-1)P(N_{\rm H:}, N_{\rm D:})\right]   \nonumber \\ 
&+& A_{\rm D} \left[(N_{\rm D:}+2)(N_{\rm D:}+1)P(N_{\rm H:}, N_{\rm D:}+2) 
- N_{\rm D:}(N_{\rm D:}-1)P(N_{\rm H:}, N_{\rm D:})\right]   \nonumber \\
&+&{(A_{\rm H}+A_{\rm D}) 
\left[(N_{\rm H:}+1)(N_{\rm D:}+1)P(N_{\rm H:}+1, N_{\rm D:}+1)
- N_{\rm H:} N_{\rm D:} P(N_{\rm H:}, N_{\rm D:}) \right]}.   
\end{eqnarray}

\noindent
Additional terms proportional to $f_H$ and $f_D$ 
are introduced in the master equations to account the Langmuir 
rejection due to D atoms.  
The derivation of the moment equations is 
very similar to the previously described 
procedure, leading to the following 
set of coupled equations for the first and 
second moments:

\begin{eqnarray*}
\frac{d \langle {N_{\rm H:}}\rangle}{dt}   &=& 
F_{\rm H} + (2 A_{\rm H} 
- W_{\rm H}-f_{\rm H}) \langle N_{\rm H:} \rangle  
- f_{\rm H} \langle N_{\rm D:} \rangle 
- 2 A_{\rm H} \langle N_{\rm H:} ^2 \rangle 
- (A_{\rm H}+ A_{\rm D}) \langle N_{\rm H:}N_{\rm D:} \rangle  \nonumber \\
\frac{d \langle {N_{\rm D:}} \rangle}{dt} &= &
F_{\rm D}  -f_{\rm D} \langle N_{\rm H:} \rangle 
+ (2 A_{\rm D}- W_{\rm D}-f_{\rm D})  
\langle N_{\rm D:} \rangle - 2 A_{\rm D} \langle N_{\rm D:}^2 \rangle  
- (A_{\rm H}+ A_{\rm D}) \langle N_{\rm H:}N_{\rm D:} \rangle \nonumber \\
\frac{d \langle {N_{\rm H:}}^2 \rangle}{dt} &=& 
F_{\rm H} + (2 F_{\rm H} + W_{\rm H} +4 A_{\rm H} 
-f_{\rm H}) \langle N_{\rm H:} \rangle  
- f_{\rm H} \langle N_{\rm D:} \rangle -
(2 W_{\rm H} + 4 A_{\rm H}+2f_{\rm H}) \langle N_{\rm H:}^2 \rangle  
\nonumber \\ 
&-& (A_{\rm H}+ A_{\rm D} + 2 f_{\rm H}) 
\langle N_{\rm H:}N_{\rm D:} \rangle 
\nonumber \\ 
\frac{d \langle {N_{\rm D:}}^2 \rangle}{dt} &=& 
F_{\rm D} - f_{\rm D} \langle N_{\rm H:} \rangle 
+ (2 F_{\rm D} + W_{\rm D} +4 A_{\rm D} 
- f_{\rm D}) \langle N_{\rm D:} \rangle - 
(2 W_{\rm D} + 4 A_{\rm D} + 2  f_{\rm D}) 
\langle N_{\rm D:}^2 \rangle  
\nonumber \\
&-& (A_{\rm H}+ A_{\rm D} + 2 f_{\rm D}) 
\langle N_{\rm H:}N_{\rm D:} \rangle  
\nonumber \\
\frac{d \langle {N_{\rm H:}N_{\rm D:}} \rangle}{dt} &=& 
F_{\rm D} <N_{\rm H:}> 
+ F_{\rm H}  \langle N_{\rm D:} \rangle  
- f_{\rm D} \langle N_{\rm H:} ^2 \rangle
- f_{\rm H} \langle N_{\rm D:}^2 \rangle
- (W_{\rm H}+W_{\rm D}
\nonumber \\
&+& A_{\rm H} + A_{\rm D} + f_{\rm H}+f_{\rm D}) 
\langle N_{\rm H:}N_{\rm D:} \rangle  
\label{equ:Mom}
\end{eqnarray*}

\noindent
{\bf This is the set of equations that is actually used in the
interstellar chemistry model, in the case where rejection
due to both H and D atoms is included}.
The production rates of 
\hh, HD and \dd on a single grain are given 
by the same equations as those reported 
previously [Eq. (\ref{equ:form})]. 
Integrating over the grain size distribution, 
we obtain the formation rates (in \ccm \sm) as well as the adsorption 
and desorption rates and the number of adsorbed hydrogen and deuterium 
atoms (in \ccm) as described in the next section. 

%

\section{Calculations and results}
\label{sec:result}

\subsection{Fixed dust temperatures}
We consider first a diffuse cloud model whose parameters are 
given in Table \ref{tab:model}. 
The chemical network comprises of 134 chemical species 
containing H, D, C, N, O, S and a typical metal "Fe"  
undergoing charge exchange reactions only in addition 
to photoionisation and electronic recombination.

We compare the results obtained via several approximations 
within the rate and the moment equations on the formation 
of \hh, HD and \dd on amorphous carbon grains (see Table~\ref{tab:grain} for their properties) for different fixed relevant 
dust temperatures (between 8 and 20K) and their link to the gas phase chemistry. 
We use the Meudon PDR code 
\citep{lepetit:02,lepetit:06} 
in which we have introduced the appropriate changes. 
The model corresponds to a slab of gas of 6.06 pc 
(corresponding to a total visual magnitude of 1) 
irradiated from both sides by the standard interstellar 
radiation field as given by 
\cite{draine:78}.


\subsubsection{Incorporation of the moment equations into the Meudon PDR code}

We couple the full gas phase network and the surface chemistry of  
\hh, HD and \dd 
in the PDR code by solving explicitely
the steady state of Eq. \ref{equ:Mom}, 
describing the evolution of the first and second moments 
of the distribution of adsorbed hydrogen and deuterium 
atoms on a single grain. 
The values of the first order moment give the average numbers 
of adsorbed H and D atoms on a single grain whereas no obvious physical 
meaning is associated to the second order moments.  
We perform the integration on the size distribution to 
get the appropriate quantities such as the
number densities of adsorbed H, D and gas phase 
atomic and molecular number densities. 
First, we have checked the relevance of the treatment 
by considering only \hh~formation by comparing 
with the analytical
solution obtained from the resolution of the 
two coupled equations 
\citep{lipshtat:03}.

\begin{table*}[!t]
\centering
\caption{The parameters used in the model. 
Numbers in parentheses refer to powers of ten. N$_\textrm{H}$ is the total column density of protons (N(H) + 2 N(H$_2$)) and A$_\textrm{v}$ is the visual extinction.
}
\label{tab:model}
\begin{tabular}{|lc||lc|}
\hline
Gas parameter   &   & Dust  parameter &    \\
\hline
n$_H$ (cm$^{-3}$) &  100 & $a_{min}$ (cm)    & 3(-7)   \\
$T_{gas}$  &  70 & $a_{max}$ (cm) & 3(-5)  \\
$\zeta$ (s$^{-1}$) & 5(-17) & 
$q$  & 3.5   \\
$\chi$ (Draine unit) & 1 & 
material   & amorphous carbon   \\
D/H  & 1.5(-5) & $\gamma$  & 1  \\
C/H  & 1.34(-4) &  $G$ & 0.01   \\
O/H  & 3.19(-4) & total visual magnitude      & 1\\
N/H  & 7.5(-5) & N$_\textrm{H}$ / A$_\textrm{v}$ & 1.871(21) \scm /magnitude \\
S/H  & 1.86(-5)   &   T$_{dust}$  &   fixed [8 - 20K] \\
Fe/H & 1.5(-8) &    &   \\

\hline
\end{tabular}
\end{table*}

\subsubsection{Formation of \hh~ and HD on grains}

We have examined the assumptions described above, 
using both the moment equations and the rate equations.
More specifically, we compared three models:
a model that includes no Langmuir rejection, 
a model that includes rejection only due to adsorbed
H atoms and a model that includes rejection due to both
H and D atoms adsorbed on grains.
The three models are listed in table
\ref{tab:hyp}.

\begin{table*}[!t]
\centering
\caption[]{The three models simulated using the rate equations
and the moment equations and their legends.}
\label{tab:hyp}
$$ 
\begin{tabular}{|lllll|}
\hline
\noalign{\smallskip}
Model    &  rate/moment &   No rejection  &  
Rejection for H only & Rejection for H and D   \\
\noalign{\smallskip}
\hline
\noalign{\smallskip}
RA   &  Rate equations &  yes  & -    &    -\\
RB  &  Rate equations & -  & yes &   - \\
RC &  Rate equations &  -& - &   yes\\
MA &  Moment system &  yes & -  &   - \\
MB  &  Moment system &  - & yes &   - \\
MC  &  Moment system &-  & - &   yes \\
\noalign{\smallskip}
\hline
\end{tabular}
$$ 
\end{table*}

\begin{figure}[h]
\centering
\begin{center}
\includegraphics[width=1\linewidth]{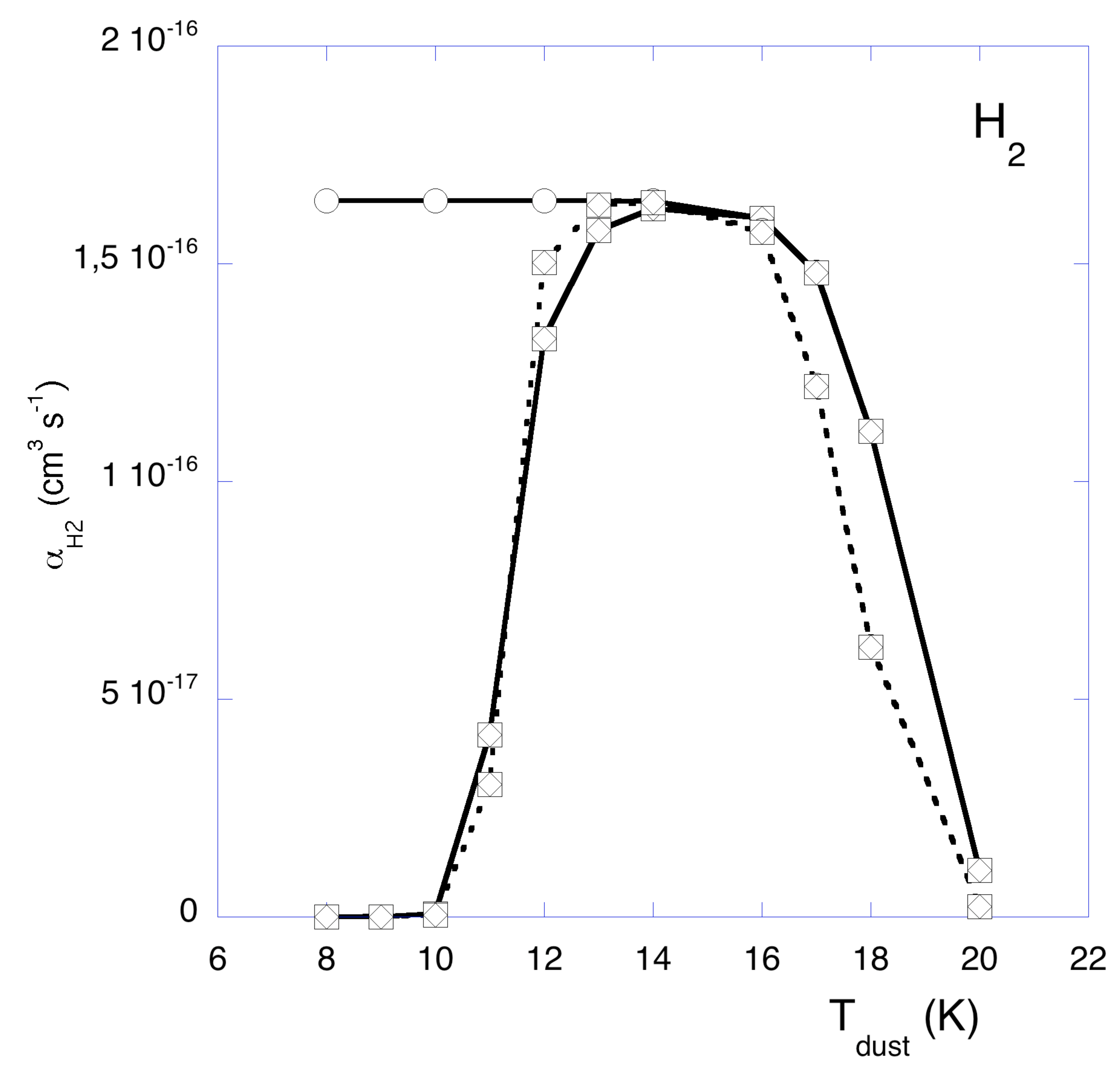}
\caption{The formation rate 
$\alpha$(H$_2$) (\cpm \sm)
of \hh ~molecules on grains, 
obtained from the rate equations (solid line) and from the moment equations (dashed lines),
where the grain sizes 
follow the MRN size distribution. 
Results obtained from the three models of Table 3 are presented:
Model A which includes no rejection (open circles), model B which includes rejection only due to
adsorbed H atoms (open squares) and model C which includes rejection due to both adsorbed H and
adsorbed D (open diamonds).
}
\label{fh2}
\end{center}
\end{figure}

\begin{figure}[h]
\begin{center}
\includegraphics[width=1\linewidth]{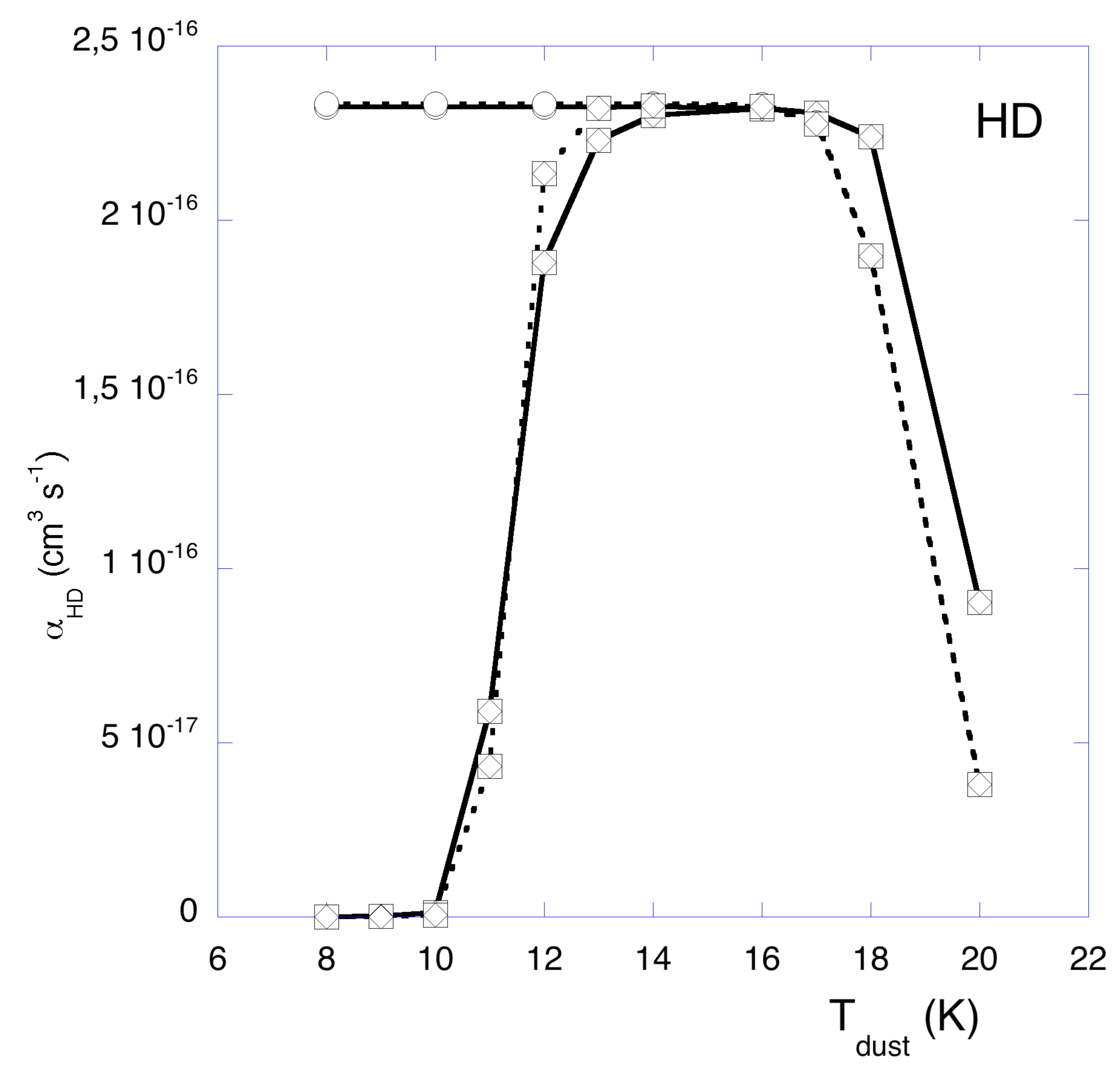}
\caption{The formation rate 
$\alpha$(HD) (\cpm \sm)
of HD molecules on grains, 
obtained from the rate equations and the moment equations,
where the grain sizes 
follow the MRN size distribution. Same conventions as in Fig. \ref{fh2}
}
\label{fhd}
\end{center}
\end{figure}

We will present the results in the form of the standard
rate law for the production of molecular hydrogen 
in interstellar clouds.
In this rate law, the production rate $R_{\rm H_2}$
(cm$^{-3}$ s$^{-1}$)
is expressed by

\begin{equation}
R_{\rm H_2} = \alpha({\rm H_2}) n({\rm H}) n_{\rm H},
\label{eq:rate_lawh2}
\end{equation}

\noindent
where $n_{\rm H}$ (cm$^{-3}$)
is the total density of H nuclei, in atomic
and molecular forms.
It can be approximated by
$n_{\rm H}= n({\rm H}) + 2 n({\rm H_2})$.
The parameter $\alpha({\rm H_2})$
(cm$^{3}$ s$^{-1}$)
is the effective rate coefficient.
In the case of diffuse clouds, this rate coefficient
is often crudely approximated by
$\alpha({\rm H_2}) = 5 \times 10^{-17} \sqrt{T_{\rm gas}/300}$.
The formation rate of HD molecules can be expressed in
a similar way, by

\begin{equation}
R_{\rm HD} = \alpha({\rm HD}) n({\rm D}) n_{\rm H},
\label{eq:rate_lawhD}
\end{equation}

\noindent
The parameter $\alpha({\rm HD})$
(cm$^{3}$ s$^{-1}$)
is the effective rate coefficient
for HD formation on dust grains.

In Fig. \ref{fh2} we present the rate
coefficient
$\alpha({\rm H_2})$
for the formation of H$_2$ molecules on grains
and
in Fig. \ref{fhd}
we present the rate coefficient
$\alpha({\rm HD})$
for the formation of HD molecules.
These rate coefficients are plotted
versus the assumed fixed temperature of the grains
for the three models specified above,
simulated by both the rate equations and
the moment equations.
The conditions used in the simulations
are those  
obtained at the edge of the modelled cloud, 
i.e. A$_\textrm{V}$ = 0. 
When the Langmuir rejection effect is taken into account,
the formation of molecular hydrogen on grains is efficient 
within a narrow window of the grain temperatures,
in agreement with previous studies 
\citep{Katz:99,lipshtat:04}.
At higher temperatures, the hydrogen atoms do not
stay long enough on the grain to encounter each
other and form molecules.
At lower temperatures, the grain surface is saturated by
nearly immobile hydrogen atoms, and the reaction rate
is very low. 
These adsorbed atoms reject new atoms that impinge on 
the surface and prevent the formation of additional
layers of hydrogen atoms on the surface.
When the Langmuir rejection term is removed from the
equations, the range of high efficient recombination is
extended towards lower temperatures.
However, in this case the model enables the accumulation
of many layers of adsorbed hydrogen atoms on the surface.
Such accumulation is unphysical. Furthermore, 
under such circumstances, the formation of molecular
hydrogen is not mediated by diffusion, and thus the
model itself is not valid.
Laboratory experiments provide strong evidence for 
Langmuir rejection
\citep{Katz:99}.
Therefore, the standard forms of the rate equations and
of the moment equations which we incorporate in the PDR
code are those that include the Langmuir rejection term.
In presence of this term, we observe that
the high efficiency window is somewhat larger for HD 
formation than for H$_2$ formation.
This is due to the somewhat higher desorption barrier for 
D atoms than for H atoms. 
The maximum value of the formation rate of \hh
shown in Fig. 1 is in excellent agreement with
the "effective" formation rate 
$\alpha_{H_2}$  (\cpm \sm), 
expressed by
\citep{lebourlot:95} 

\begin{equation}
\alpha_{H_2} = 
%
%
1.4 \cdot \frac{3 \cdot m_{\rm H} \cdot G }
{4 \cdot \rho \cdot \sqrt{a_{min}a_{max}}}
\upsilon_{\rm H}  = 1.62 \times 10^{-16} n_{\rm H}, 
\end{equation}

\noindent
with the chosen parameters. 
The corresponding value for HD 
is 
$\alpha_{HD} = \sqrt{2} \alpha_{H_2} = 2.3 \times 10^{-16} n_{\rm H}$ 
(\cpm \sm), 
as seen on Fig. \ref{fhd}. 
The rate equations and the 
moment equations are almost identical at 
the maximum when desorption 
effects are not significant. 
However, significant discrepancies appear 
at the edges. 
Rejection effects play a significant role at the lowest 
temperatures of the grains, which may not be very relevant 
for astrophysical purposes. 
We display in Table \ref{tab:cd} 
the resulting column densities of gas phase 
H, \hh, D and HD when integration of the abundance 
densities is performed over the width of the cloud. 
We have chosen to display only the results of  
model C which are the most accurate as rejection 
effects are included both for H and D impinging atoms.

\begin{table*}[!t]
\centering
\caption[]{Column densities (in \pscm) 
of H, H$_2$, D and HD
and the molecular fraction f
for model C, obtained from the rate equations (RC) and from the moment
equations (MC) for different grain temperatures. 
Numbers in parentheses refer to powers of ten.
}
\label{tab:cd}
$$ 
\begin{tabular}{|lllllll|}
\hline
\noalign{\smallskip}
Model    & $T_{dust}$& H &   \rm {H$_2$} & f & D & HD   \\
\noalign{\smallskip}
\hline
\noalign{\smallskip}
RC   &  8&  1.87(21)  & 1.60(14)  &1.7(-7)  & 2.76(16) &2.64(10) \\
MC  &       8      & 1.87(21)  & 1.60(14)& 1.7(-7)  &  2.81(16)& 6.26(9) \\
RC   &  10&  1.77(21)  & 5.30(19) & 5.65(-2) &  1.74(16)& 1.07(16) \\
MC  &     10    &  1.84(21) & 1.62(19)& 1.73(-2) &2.28(16)& 5.25(15) \\
RC   &  12&  6.65(19)  & 9.02(20)& 0.96 & 1.26(16)  & 1.55(16) \\
MC  &    12    &  6.33(19) & 9.04(20)& 0.97&  1.26(16) &  1.55(16)\\
RC   &  14&  6.29(19) & 9.04e(20)& 0.97&1.26(16) &1.55(16)\\
MC  &   14     &  6.29(19) & 9.04(20)&0.97&   1.26(16) &  1.55(16) \\
RC   &  16&  6.95(19) & 9.01(20) &0.96 &  1.25(16)&1.56(16)\\
MC  &   16     & 8.87(19) & 8.91(20) & 0.95 & 1.21(16)&  1.60(16) \\
RC   &  18&  1.76(20)& 8.48(20)  & 0.91 &1.07(16)&1.74(16)\\
MC  &   18    &  3.43(20)&  7.64(20) &  0.82 & 8.91(15& 1.91(16) \\
RC   &  20&  9.60(20)& 4.55(20) & 0.49  &7.58(15) & 2.05(16) \\
MC  &   20     & 1.52(21) & 1.77(20) & 0.19  &1.12(16)& 1.69(16)\\
\noalign{\smallskip}
\hline
\end{tabular}
$$ 
\end{table*}

\noindent
The column densities of \hh~display significant 
variations with the assumed dust temperature as reflected by the values of the 
molecular fraction defined by 
$f = {{2 N(H_2)}/{[N(H) + 2 N(H_2)]}}$.
The resulting column densities of HD become less sensitive 
as soon as some molecular hydrogen is present.
This is due to the fact that in presence of H$_2$
the 
formation of HD occurs mainly via gas phase processes. 
The formation of HD on grains 
is more efficient than in the gas phase
only at the edge where no \hh ~is yet formed. 
We discuss this in more detail in the next section where we introduce a variable
dust temperature as a function of the visual magnitude.
%
\subsection{$A_{\rm v}$ dependent dust temperatures} 
%
\subsubsection{\bf {Homogeneous temperature distribution}}
\label{subsec:AVdependenttemp}
It is well recognised that the grain temperatures decrease when the strength of the 
impinging radiation field decreases for increasing visual magnitude. We introduce 
such a dependence by using 
 the simple analytic formula
presented by \cite{hollenbach:1991}, 
which provides the grain temperature as a function of the
strength of the radiation field and the visual magnitude. 
As our model involves a slab of gas, irradiated from both sides,
we extend the formula of \cite{hollenbach:1991} by introducing 
$[\chi_{\rm left} \exp(-1.8 A_{\rm v}) + \chi_{\rm right}  exp(-1.8(A_{\rm v}^{\rm tot}-A_{\rm v})]$
for the A$_{\rm v}$ dependence, where $\chi_{\rm left}$ and $\chi_{\rm right}$ 
are the strengths of the radiation field on both sides, 
expressed in Draine units, and $A_{\rm v}^{\rm tot}$ is the total visual magnitude of the cloud.
In this approximation, grains of all sizes exhibit the same average temperature. We consider possible
fluctuations effects in \ref{subsec:fluctuations}

\begin{table}[h]
\caption[]{Physical conditions relevant to a dense PDR.}
\label{tab:pdr}
\begin{tabular}{cc}
\hline
Physical quantity    &  value \\
\hline
\noalign{\smallskip}
$n_{\rm H}$  &  10$^4$ \ccm\\
$A_{\rm v}^{\rm tot}$ &  10   \\
size &   0.61 pc \\
 $\chi_{\rm left} $ &  100  \\
 $\chi_{\rm right}$ &  1  \\
 $T_{\rm gas}$   &   30K  \\
 $T_{\rm dust}$   &  $A_\textrm{v}$ dependent  \\
 size distribution of grains &  same as in Table \ref{tab:model} \\

\noalign{\smallskip}
\hline
\end{tabular}
\end{table}

In order to span a significant range of dust temperatures, we consider  
a dense photodissociation region (PDR) with a radiation field strength 
($\chi_{\rm left} $) of 100 impinging on the left side and the standard radiation 
field ($\chi_{\rm right}$) impinging
on the right side. The parameters of the model are displayed in 
Table \ref{tab:pdr}. As we want here to discuss the different effects 
of the rate versus moment equations systems, we keep all the physical parameters identical.

\begin{figure}[h]
\begin{center}
\includegraphics[width=1\linewidth]{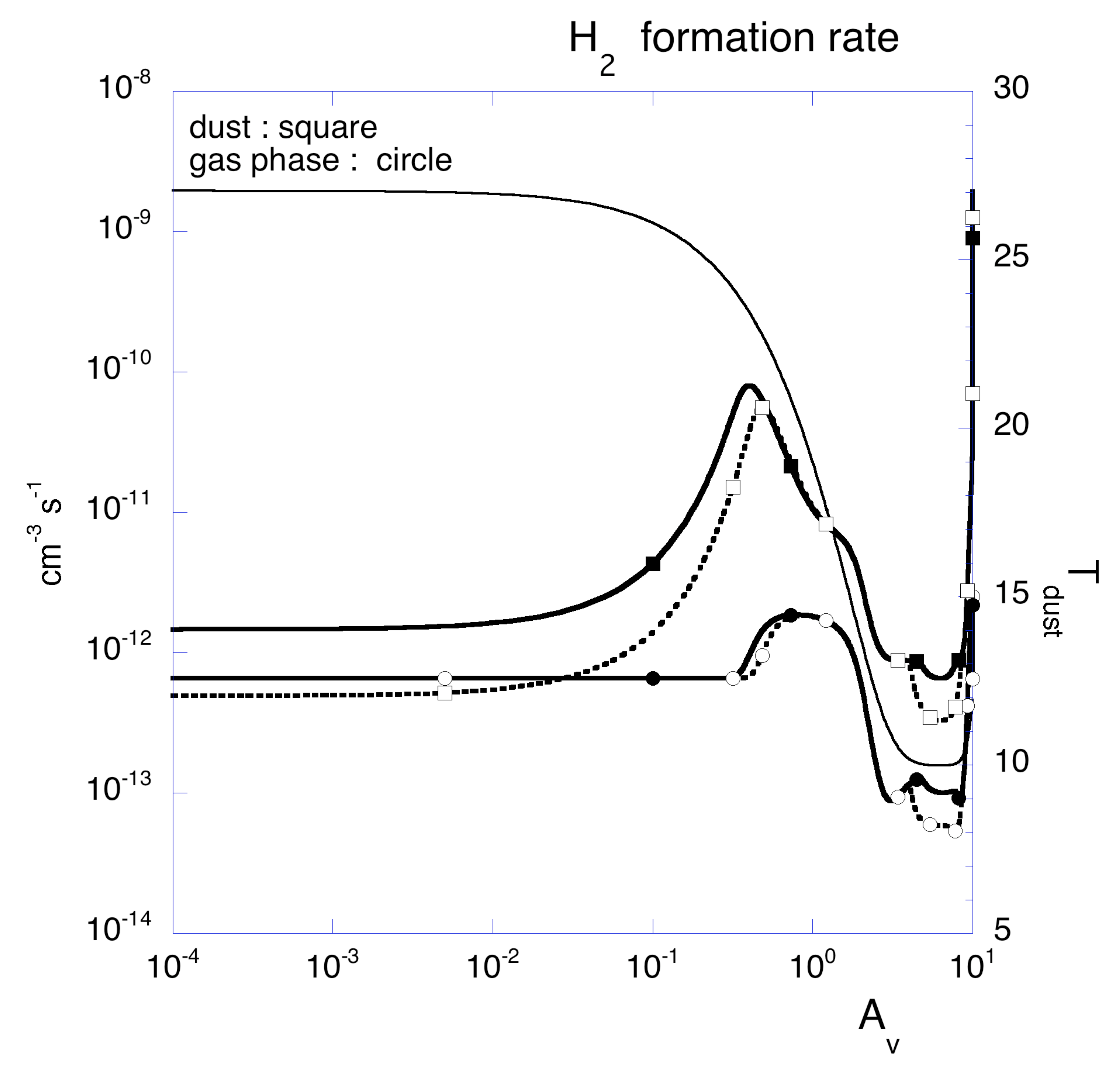}
\caption{Comparison between gas phase formation (circles) and grain-surface 
formation (squares) rates of H$_2$ (in \ccm \sm),
obtained from the rate equations (solid lines and full symbols)
and from the moment equations (dotted lines and open symbols).
The thin solid line shows the grain temperature
(the temperature scale is on the right hand side) 
}
\label{fh2gg}
\end{center}
\end{figure}

\begin{table*}[!t]
\centering
\caption[]{Column densities in \pscm ~and molecular fraction f for the dense PDR models. 
Numbers in parentheses refer to power of ten.
}
\label{tab:cdn4}
$$ 
\begin{tabular}{|llllll|llll}
\hline
\noalign{\smallskip}
Model   & H &   \rm {H$_2$} & f & D & HD  & C & CO  & CH \\
\noalign{\smallskip}
\hline
\noalign{\smallskip}
RC   &   2.29(21)  & 8.21(21)  &0.88 & 3.29(16) &2.28(17) & 7.74(17) &  1.18(18) & 9.31(14)\\
MC  &       6.35(21)  & 6.18(21)& 0.66  & 3.25(16)& 2.46(17)& 5.1(17) & 1.10(18) & 1.02(15) \\
\noalign{\smallskip}
\hline
\end{tabular}
$$ 
\end{table*}

\begin{figure}[h]
\begin{center}
\includegraphics[width=1\linewidth]{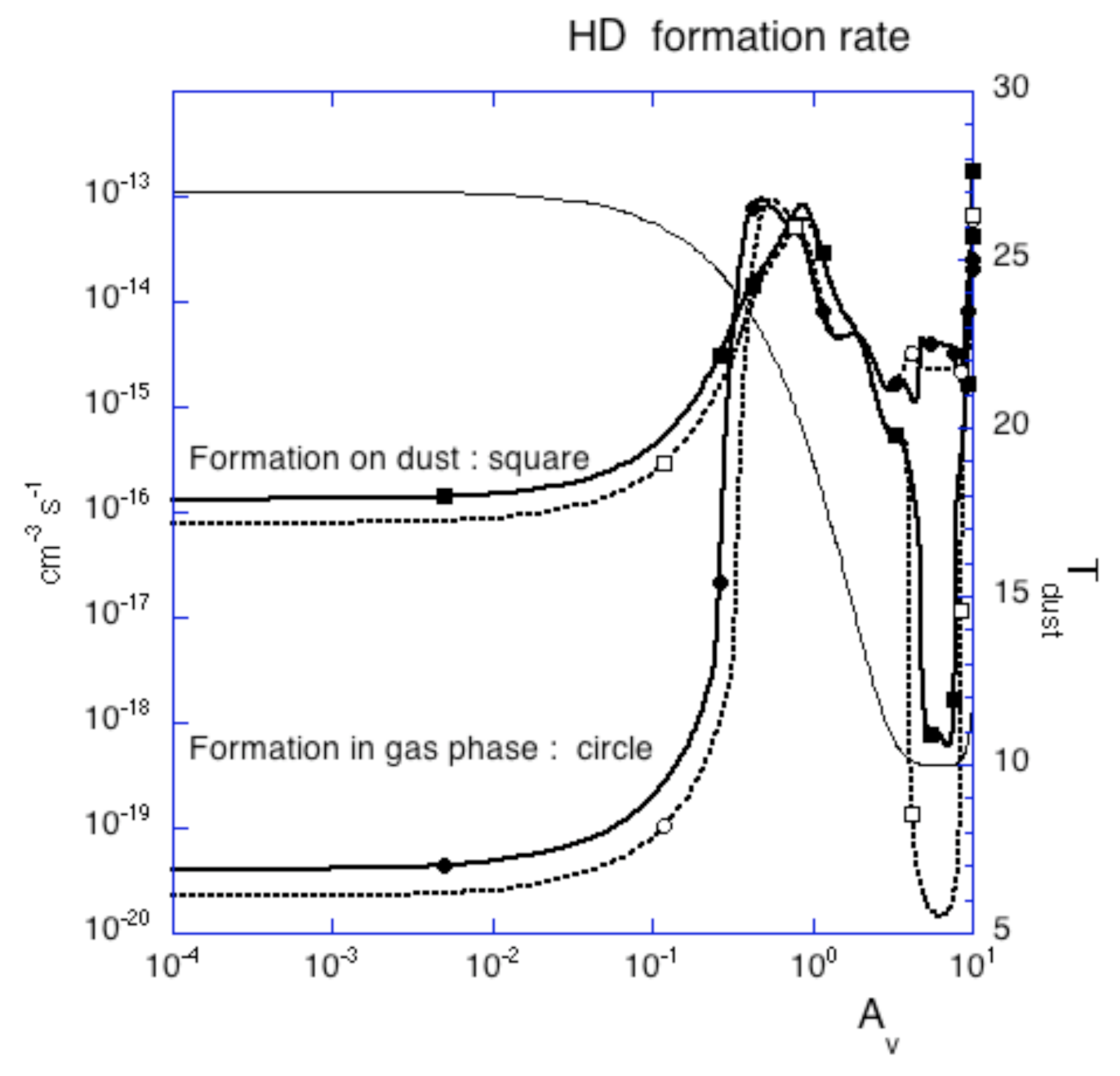}
\caption{Comparison between gas phase formation (circles) and grain-surface formation
(squares) 
rates of HD (in \ccm \sm) 
obtained from the rate equations (solid lines and full symbols)
and from the moment equations (dotted lines and open symbols). 
The thin solid line shows the grain temperature
(the temperature scale is on the right hand side) 
}
\label{fhdgg}
\end{center}
\end{figure}

We display in Figures \ref{fh2gg} and  \ref{fhdgg} the dust temperature 
variation together with the formation rates through gas phase and 
grain surface reactions of  \hh ~and HD obtained with the RC and MC 
treatments as these are expected to be the most physical within the 
rate equation and moment equation formalisms.
The dust temperature profile spans a range of values between about 27K 
on the left side, reaching a minimum of 10K in the shielded region at 
$A_\textrm{v}$  between 5-7 , and reaching a value of about 10K on the 
right edge. This range corresponds to very low to high efficiencies of 
formation of H2 and HD via grain surface reactions;  
As far as \hh~ formation is concerned, we see that the difference between 
rate and moment equation formalisms is significant at the left side of the 
cloud where the dust temperature reaches a value of 27K, corresponding to 
a situation where H atoms desorb efficiently from the surface. 
The results obtained via the rate equations are larger by a factor of about 2. 
In this particular case, this leads to a competition between gas phase and dust processes. 
However, as soon as the dust temperature reaches a value of about 25, 
the formation efficiency on dust becomes preponderant. Both treatments 
are equivalent for a range of temperature between 20 and 11K as already 
shown on Figure \ref{fh2}. 
Significant discrepancies occur again for the range of visual magnitude 
5-8 where the dust temperature reaches a value of 10K.

These behaviours have a direct impact on the HD formation mechanisms, as displayed on Figure \ref{fhdgg}.  As long as no significant formation of \hh~occurs, 
HD is mainly formed via dust processes. However, as soon as \hh~ is present, formation of HD via gas phase is much more efficient.  In the shielded region where $A_\textrm{v}$ reaches a value of a few and the dust temperature is about 10K, significant variations occur due to rejection effects. Table \ref{tab:cdn4} displays the resulting column densities. Whereas significant differences arise for the column densities of H and \hh, the values relevant to the other species are quite similar. 

\subsubsection{Temperature fluctuations effects}
\label{subsec:fluctuations}

The consideration of a homogeneous distribution of dust temperatures as a function of A$_\textrm{v}$ is reasonable for low impinging radiation fields and grains of radii larger than $1.0 \times 10^{-6}$ cm, 
\citep{Horn2007}. 
For grains of radii smaller than $\sim$  $1.0 \times 10^{-6}$ cm, the absorption of a single UV photon may cause a temperature spike that heats up the
grains by over 20 K \citep{Horn2007}. Such a spike is likely to lead to the immediate desorption of all the H and D atoms adsorbed on the grain surface. 
This conclusion is also derived from detailed kinetic Monte Carlo simulations of H$_2$ formation on stochastically heated olivine grains \citep{Cuppen2006MNRAS,Cuppen2006PNAS}. 
To quantify the effect of temperature fluctuations on small grains, we have run similar models than those presented previously, except that we cut the size distribution of grains at the minimum value of $1.0 \times 10^{-6}$ cm. We present in Table \ref{tab:cdn4fluc} the corresponding column densities. In this way, we do not take into account any production of molecular hydrogen on grains of radii smaller than $1.0 \times 10^{-6}$ cm both in rate equations and systems of moments approach. 
\begin{table}[h!]
\centering
\caption[]{Column densities in \pscm ~and molecular fraction f for the dense PDR models in which surface chemistry is cut  for very small grains (r $<$ $1.0 \times 10^{-6}$ cm). Numbers in parentheses refer to power of ten.}
\label{tab:cdn4fluc}
\begin{tabular}{lllllllll}
Model &  H   &    H$_2$   &  f   &   D   &  HD  &  C  & CO  &  CH \\ 
RC'    & 5.88(21) & 6.41(21) &  0.69   & 3.52(16)  &   2.43(17)  &  8.45(17) & 1.11(18) & 1.02(15) \\  
MC'    & 8.40(21) & 5.16(21) &  0.55   & 3.47(16)  &   2.46(17)  &  9.27(17) & 1.02(18) & 1.06(15) \\
\end{tabular}
\end{table}
  In these models, computed column densities of H$_2$ and HD  are smaller than those obtained with the MRN distribution extended to $3.0 \times 10^{-7}$ cm (cf Table \ref{tab:cdn4fluc}) due to a smaller integrated grain cross section value. We find that
  rate and moment equations give  still significantly different values of the total column densities of H$_2$ 
  even for these larger grains. These differences arise essentially from the regions when T$_{dust}$   is larger than about 18K, i.e. at the edges of the cloud.\\
 The present results give an upper limit of the effect of temperature fluctuations as these can only result in a decrease of molecule formation at the surface of grains. In addition, these effects will not occur at A$_\textrm{v}$ values larger than $\sim$ 1  and  smaller than $\sim$ 9 as the radiation field is then considerably reduced.
%
\section{Summary}
\label{sec:conclusion}

We have incorporated the moment equations for the formation
of H$_2$ and its deuterated versions into the Meudon PDR 
code, examined their applicability and relevance
and compared the results with those obtained from
rate equations incorporated into the same code
under identical conditions. 
Previous analysis has shown
that as long as the temperatures of the dust grains
are not too high, the reaction rates obtained from the 
moment equations coincide with those obtained from the 
rate equations. At grain temperatures around 18 K or higher,
on the high-temperature end of the efficiency window 
there are significant differences, where the rate equations
over-estimate the formation rates of H$_2$ and HD.
These deviations are mainly due to the very small grains,
on which the population sizes of adsorbed H and D atoms
exhibit large fluctuations.
Under such conditions it is important to use the 
moment equations rather than the rate equations.
For low grain temperatures, below 12 K or so,
we find that the Langmuir rejection term makes a crucial
difference. In this range, the rejection term is required 
in order to prevent the freezing of multi-layers of 
hydrogen atoms on the grains. Such freezing is unphysical.
Here we have derived for the first time a set of moment 
equations in which the Langmuir rejection term is incorporated.
The comparison was made with rate equations,
in which a suitable rejection term was also incorporated.
For very low temperatures, where the grains are covered by
a layer of H and D atoms and the rejection term plays a
major role, we also find a discrepancy between the
formation rates of H$_2$ and HD obtained from the
rate equations and the moment equations. 
Under these conditions, deep inside a molecular cloud,
the formation rate and dissociation rate of H$_2$
molecules are low, while HD molecules form mainly by
gas phase reactions.
These are conditions under which both the rate equations
and the moment equations are accurate. 
The observed discrepancy is due to somewhat different
physical consitions that emerge as a result of the
discrepancy in warmer regions, where the
moment equations are accurate and the rate equations are not.
Therefore, the moment equation results are those one
can rely on also in the cold regions.

We conclude that the moment equations provide an 
efficient method for the evaluation of molecular hydrogen
formation in interstellar chemistry codes. 
The equations provide accurate results for the reaction
rates on both large and small grains.
The equations are easy to construct and are
efficient in terms of computational resources.
They can be easily coupled to the rate equations
of gas-phase chemistry.
Since the moment equations are linear, their stability 
and convergence properties
are often superior even in comparison to the rate equations.

The present study has been achieved by considering that molecular hydrogen
can be formed only through diffusion of adsorbed atoms, with the strong constraints of 
present experimental knowledge. Other formation mechanisms may be at work, as those suggested
by \cite{cazaux:04} and \cite{habart:04} in their interpretation of observations of warm molecular hydrogen.

The method can now be extended to more complex reaction networks
on grains.
The network that
involves H, O and CO molecules, from which ice mantles
 containing H$_2$O, CO$_2$ and CH$_3$OH 
are formed by successive hydrogenation and oxydation reactions,
is promising \citep{barzel:07}.
Unlike the H and D atoms, these heavier atoms and molecules
are more strongly bound to the surface and to each other
and do not exhibit the Langmuir rejection property.
Therefore, in the construction of moment equations for
more complex networks one only needs to maintain
the rejection terms for H and D, presented above
and there is no need to incorporate such terms for
any other atomic or molecular species.
Recent studies have shown that CH$_3$OH molecules do not form in the
gas phase as efficiently as previously expected.
The approach presented here will enable to evaluate the formation rate
of CH$_3$OH on
dust grains 
under different physical conditions, taking into account the contributions
of grains of different sizes.

\begin{acknowledgements}

This work was supported in part by the France-Israel 
High Council for Science and Technology Research via contract 07 AST F.

\end{acknowledgements}

\bibliographystyle{aa}
\bibliography{LePetit09}

\end{document}